\documentclass[a4paper,twocolumn,fleqn]{article}

\usepackage{txfonts}               
\usepackage{bm}                  
\usepackage{graphicx}     
\usepackage{pfr_arXiv}                   
\usepackage{url}

\begin{document}

\title{GPU-Portable Real-Space Density Functional Theory Implementation on Unified-Memory Architectures}


\author{Atsushi~M.~ITO\sup{1,2}}

\affiliation{
  \sup{1}National Institute for Fusion Science, National Institutes of Natural Sciences, Oroshi-cho 322-6, Toki 509-5292, Japan.\\
  \sup{2}Graduate Institute for Advanced Studies, SOKENDAI, Oroshi-cho 322-6, Toki 509-5292, Japan.}

\date{(Received XX Month YYYY / Accepted XX Month YYYY)}

\email{ito.atsushi@nifs.ac.jp}

\begin{abstract}
We present a GPU-portable implementation of a real-space density functional theory (DFT) code ``QUMASUN'' and benchmark it on the new Plasma Simulator featuring Intel Xeon 6980P CPUs, and AMD MI300A GPUs. Additional tests were performed on an NVIDIA GH200 GPU. In particular MI300A supports unified memory and GH200 supports coherent memory interconnect, simplifying GPU porting.
A lightweight C++ lambda-based layer enables CPU, CUDA, and HIP execution without OpenMP/OpenACC preprocessor directives. For diamond (216 atoms) and tungsten (128 atoms) systems, MI300A and GH200 achieve 2.0-2.8 $\times$  and 2.3-2.4 $\times$ speedups over a 256-core Xeon node. The compute-bound kernels, which are fast Fourier transforms (FFT), dense matrix-matrix multiplications (GEMM) and eigenvalue solver, show substantial acceleration on both GPUs, indicating that the present GPU-portable approach can benefit a wide range of plasma-fusion simulation codes beyond DFT.
\end{abstract}



\maketitle  


Understanding plasma-surface interaction phenomena at the microscopic level requires the first principle molecular simulation techniques. In our recent studies\cite{1,2}, we have investigated the neutralization of energetic ions incident from plasmas onto solid surfaces using Ehrenfest molecular dynamics (Ehrenfest MD), where nuclear motion is treated classically and the time evolution of valence electrons is computed within time-dependent density functional theory (TDDFT)\cite{EhrenfestMD}. For describing such excitation processes and non-adiabatic processes, the real-space density functional theory (RS-DFT), which represents wave functions on a real-space grid without employing basis functions, is particularly advantageous and provides a straightforward path to TDDFT extensions.

In this work, we extended our QUMASUN code, which has been developed to perform RS-DFT, RS-TDDFT, and Ehrenfest MD, to support execution on Graphics Processing Units (GPUs). RS-DFT requires computationally intensive components such as fast Fourier transforms (FFT), dense matrix-matrix multiplications, and generalized eigenvalue solvers. These kernels map efficiently onto GPUs. Motivated by these developments, we implemented GPU support targeting unified-memory-capable AMD MI300A and NVIDIA GH200 devices and evaluated their computational performance. The GPU architectures equipped with unified memory increasingly mitigate data-transfer bottlenecks between CPUs and GPUs. The MI300A is deployed in the new Plasma Simulator, a joint supercomputing system operated by National Institutes for Quantum Science and Technology (QST) and National Institute for Fusion Science (NIFS) since July 2025.


We benchmark only the self-consistent field (SCF) scheme part to solve an electronic ground-state, which is the most computationally demanding. As a RS-DFT code, QUMASUN performs four major computational tasks in each SCF iteration: (1) FFT used to apply the Laplacian to electronic orbitals; (2) dense matrix-matrix multiplications required to calculate Hamiltonian matrix $H$ and overlap matrix $S$ in the locally optimal block preconditioned conjugate gradient (LOBPCG) method\cite{LOBPCG}; (3) a generalized Hermitian eigenvalue solver also used in LOBPCG; and (4) the nonlocal operator of the pseudo-potential. Among these, (1)-(3) are compute-bound, whereas (4) is memory-bandwidth-bound. FFT operations are performed using FFTW (Intel MKL), cuFFT, or rocFFT depending on the system, and all electronic orbitals are processed simultaneously using batched routine. Dense matrix-matrix multiplications are executed with Intel MKL, cuBLAS, or rocBLAS.

The generalized Hermitian eigenvalue problem in LOBPCG is $HX=\lambda SX$, where $H$ and $S$ are Hermitian matrices. LAPACK provides ZHEGVD for generalized eigenvalue solver with divide-and-conquer routine; however, ScaLAPACK does not provide routine for this task which would be called PZHEGVD. By applying a Cholesky factorization $S=LL^{\dagger}$, the problem is transformed into a standard Hermitian eigenvalue problem 
\begin{equation}
KZ = \lambda Z, \quad K=L^{-1}H(L^{\dagger})^{-1}, \quad Z=L^{\dagger} X
\end{equation}
which is solved by (P)ZHEEVD. The transformation $H \rightarrow K$ is normally handled by (P)ZHEGST, but the same operation can be performed by two calls to (P)ZTRSM, which solves $Y$ from $AY=B$. This TRSM-based method was found to accelerate the transformation not only in ScaLAPACK on CPUs but also in cuSolver and rocSolver on GPUs as shown later.

For basic loop processing other than the operations described above, we ported the code to GPUs using C++ lambda expressions. No offloading directives such as OpenMP or OpenACC were used. Instead, loop bodies written as lambda expressions are transformed at compile time into either (i) ordinary CPU loop constructs or (ii) CUDA/HIP GPU kernels, depending on the compilation target. This mechanism acts as a lightweight, in-house alternative to Kokkos\cite{Kokkos}.

A simplified example for a one-dimensional loop is shown below. In a header file, we prepare the following wrapper:
\begin{verbatim}
#ifdef USE_GPU
#define GY_LAMBDA   [=]__device__
template <class FUNC>
__global__ void knl_1d(int N, FUNC func) {
  const int i = threadIdx.x
      + blockDim.x * blockIdx.x;
  if (i >= N) return; func(i); 
}
template <class FUNC>
void For(int N, FUNC func) {
  knl_1d<<<(N + Nth - 1)/Nth, Nth>>>(N, func);
}
#else   //CPU code//
#define GY_LAMBDA   [&]
template <class FUNC>
void For(int N, FUNC func) {
  for (int i = 0; i < N; ++i) func(i);
}
#endif
\end{verbatim}
where \verb*`Nth` is the number of threads per GPU block. The macro \verb*`GY_LAMBDA` defines the lambda capture mode.
With this wrapper included, a one-dimensional loop can be written simply as:
\begin{verbatim}
For(N, GY_LAMBDA(int i) {
  dest[i] = src[i] * a + b;
});
\end{verbatim}
When compiled for CPUs, this call expands into a plain for-loop, whereas on GPUs the same code is transformed into a kernel launch with a simple one-dimensional thread decomposition. The same mechanism naturally extends to multidimensional loops. Reduction operations are implemented using warp-shuffle and shared-memory techniques, with reductions confined within a single block.

More complex components, such as the nonlocal operator of pseudo-potential, are implemented with dedicated GPU kernels. Because the CUDA and HIP APIs are highly similar, portability across GPU architectures is greatly improved. In practice, QUMASUN required only minimal implementation differences between these environments.


Benchmarks were performed on a single Intel Xeon 6980P node (256 cores, FP64 peak 16 TFLOPS) and on a single AMD MI300A GPU (FP64 peak 122 TFLOPS), which provides CPU-GPU unified memory. For comparison, we also evaluated a single NVIDIA GH200 module (FP64 peak $\approx$67 TFLOPS), likewise equipped with chip-to-chip coherent memory. CPU runs used flat-MPI parallelization across 256 cores, while GPU runs measured the performance of a single device.


Benchmark targets are diamond and tungsten systems. The former consists of diamond lattice of a $3\times 3 \times 3$~supercell (216 C atoms), and the body-centered cubic (BCC) lattice of a $4\times 4 \times 4$ supercell (128 W atoms). The real-space grids contain $60^3$ and $80^3$ points, corresponding to plane-wave cutoffs of 86.7 Ry and 110.7 Ry, respectively. The numbers of valence orbitals are $N=436$ (C) and $N=901$ (W). Core electrons are treated via the frozen-core approximation using norm-conserving pseudo-potential, specifically the optimized norm-conserving Vanderbilt (ONCV) pseudo-potential\cite{Hamann,ONCV}. Exchange-correlation was modeled with the GGA-PBE functional\cite{GGAPBE}.

\begin{figure}[t]
\centering
\includegraphics[width=3.5cm]{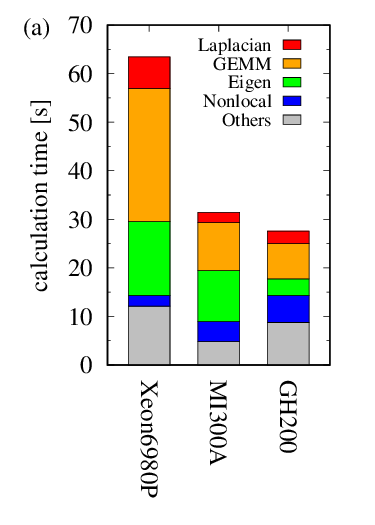}
\includegraphics[width=3.5cm]{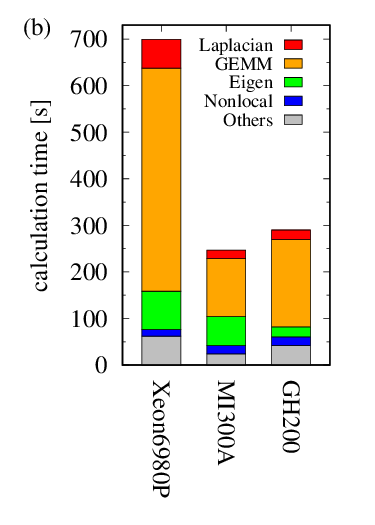}
\caption{Calculation time of SCF loop on RS-DFT for just solving 100 times eigenvalue problems for (a) diamond and (b) tungsten.}
\label{BenchDFT}
\end{figure}

Figure 1 presents the breakdown of the computational cost into the four dominant components, Laplacian (FFT), dense matrix-matrix multiplications (GEMM), eigenvalue solver, and nonlocal operator of pseudo-potential, along with ``Others'' which includes density and potential updates and MPI communication. For the diamond system, MI300A and GH200 achieve speedup factors of approximately 2.0$\times$ and 2.3$\times$ relative to a single Xeon 6980P node. For the larger tungsten system, speedups increase to 2.8$\times$ (MI300A) and 2.4$\times$ (GH200). Components (1)-(3), which are compute-bound, clearly benefit from GPU acceleration, while the nonlocal pseudo-potential component, which is memory-bound, shows modest slowdowns due to insufficient kernel optimization. Some parts of ``Others'' still fall back to host CPU loops, slightly affecting overall performance.

We note that the CPU baseline itself is strong: on the Xeon 6980P node, GEMM achieves $\approx$15 TFLOPS, close to its FP64 peak 16 TFLOPS.

\begin{figure}[t]
\centering
\includegraphics[width=7.0cm]{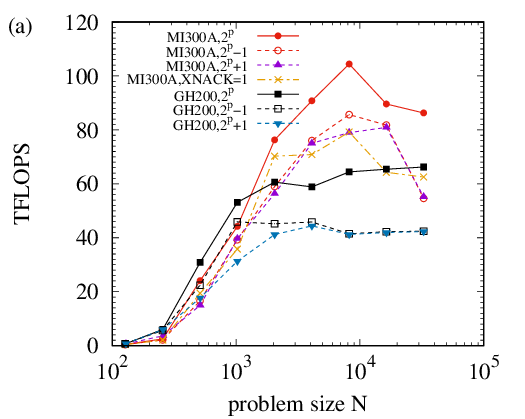}
\includegraphics[width=7.0cm]{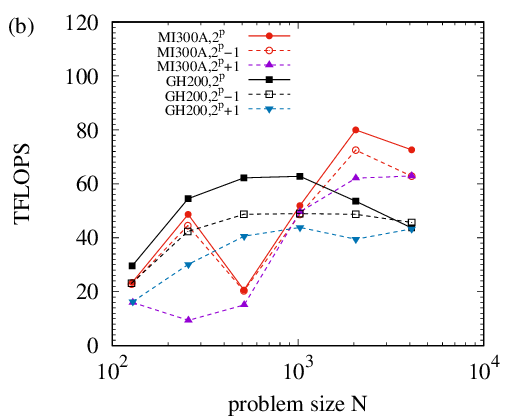}
\caption{The performance of GEMM  on Xeon 6980P, MI300A, and GH200 for (a) square matrix cases and (b) non-square matrix cases. The problem size $N = 2^p$ or $N = 2^p \pm 1$ ($p=7,8,9,\cdots$) for both cases, and $K/N = 250$ for non-square matrix cases.}
\label{BenchGEMM}

\end{figure}

To clarify the reversal of speedup trends between diamond and tungsten cases, detailed GEMM benchmarks were conducted. Figure 2(a) shows square GEMM performance for matrix size $N$, comparing powers of two $(N = 2^p)$ with nearby off-power sizes $(N = 2^p \pm 1)$. MI300A reaches $\approx 86$ \% of its theoretical peak around $N = 8192$, whereas GH200 achieves $\approx 99$ \% of its theoretical peak for square GEMM and remains above 90 \% for $N\ge 2048$. Both GPUs, however, experience a $\approx 20$ TFLOPS performance drop when matrix sizes deviate from powers of two. This explains why real-world simulation workloads cannot reach peak GEMM throughput. 

Note that MI300A provides a feature called XNACK (the X flag and non-acknowledged NACK) that allows the GPU to retry memory accesses after a page fault. However, GEMM performance decreases when XNACK is enabled. Therefore, XNACK was disabled for all other MI300A benchmarks. No issues were observed when XNACK was disabled, provided that synchronization was handled correctly.

Figure 2(b) examines rectangular GEMM for $N \times K$ matrices with aspect ratio $K/N=250$, representative of typical RS-DFT workloads. GH200 shows relatively flat performance, decreasing slightly for larger $N$, while MI300A peaks near $N\approx 2048$. For $N \ge 2048$, MI300A surpasses GH200, accounting for the higher speedup observed for the tungsten case.

\begin{figure}[t]
\centering
\includegraphics[width=7.0cm]{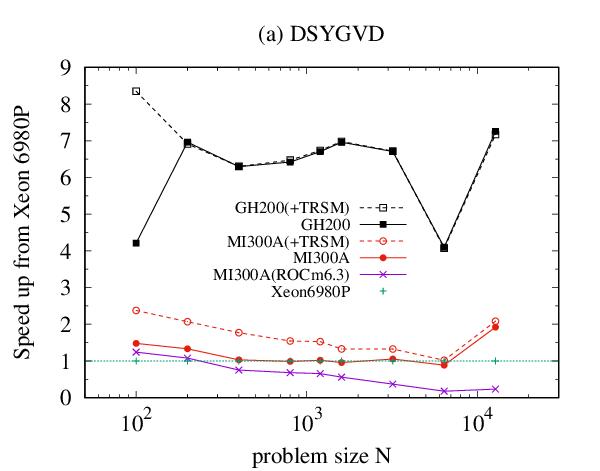}
\includegraphics[width=7.0cm]{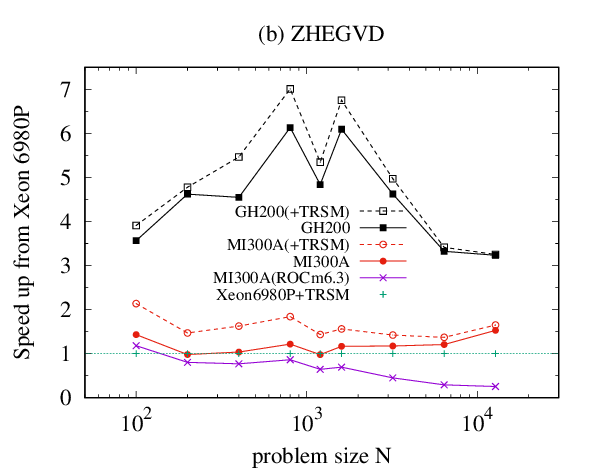}
\caption{The performance of eigenvalue solver with divide-and-conquer routine on Xeon 6980P, MI300A, and GH200 for (a) real-symmetric matrix and (b) Hermitian matrix.}
\label{BenchEigen}
\end{figure}

Eigenvalue solver performance is summarized in Fig.~3. Speedups are measured relative to ScaLAPACK (MKL) on the Xeon 6980P. For MI300A, performance depends strongly on the ROCm version: ROCm 6.3.3 falls below the CPU baseline, while ROCm 6.4.3 achieves parity. Further acceleration is obtained using the proposed transformation method based on calling TRSM twice instead of the standard xHEGST. This yields 1.5-2 $\times$ improvement on MI300A. GH200, in contrast, delivers 3-7 $\times$ speedup over the CPU baseline. For complex Hermitian problems, the TRSM-based method remains beneficial. Overall, cuSolver appears substantially better optimized than rocSolver, implying that ROCm library is still under development.

\begin{figure}[t]
\centering
\includegraphics[width=7.0cm]{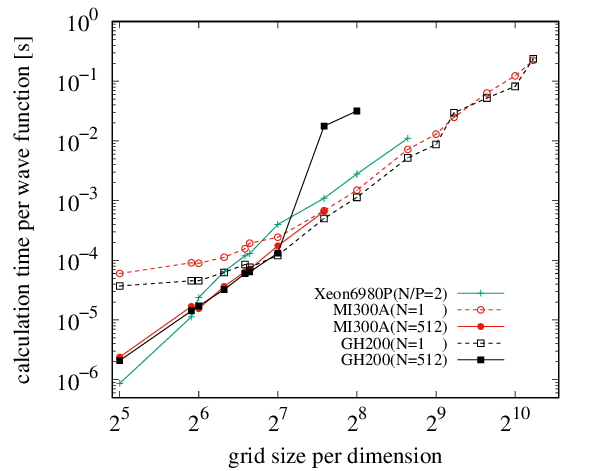}
\caption{Performance of three-dimensional FFTs for complex wave functions on the
Xeon 6980P, MI300A, and GH200. Both single FFT (N = 1) and batched FFT
(N = 512) cases were measured on the GPUs. On the Xeon 6980P, the batched
case is parallelized over P = 256 CPU cores, where each core processes
N/P = 2 wave functions. All results are shown as the execution time per
wave function.}
\label{BenchFFT}
\end{figure}

Figure 4 shows the benchmark results for the FFT, which is the primary computational cost in applying the Laplacian. On GPUs, performance degrades when repeatedly executing only a single FFT ($N = 1$), particularly when the grid size is smaller than about 128. In contrast, batch processing multiple wave functions ($N = 512$) in a single call provides significantly better performance. On the GH200, however, performance decreases once the grid size exceeds 128 because the memory required for the batched data exceeds the available GPU memory. On CPUs, good scaling is obtained by distributing $N=512$ wave functions evenly across 256 cores, where each core performs two FFTs. When the grid size is smaller than 64, the entire dataset fits within the CPU cache, which allows the CPU to outperform the GPUs in this regime.

Finally, regarding developer experience, the lambda-based portability layer allowed most CPU loops to be ported to GPUs with minimal code divergence, and unified memory on MI300A and GH200 eliminated explicit host-device transfers, simplifying maintenance. On the other hand, MI300A shows large allocation and kernel-launch overheads, requiring memory pooling and kernel fusion to achieve good performance. These architectural characteristics also contribute to the slower eigenvalue solver performance on MI300A, where many small kernels are invoked.


We implemented GPU acceleration in the RS-DFT code QUMASUN and performed SCF benchmarks on the new Plasma Simulator. We further propose the TRSM-based acceleration method of generalized eigenvalue solvers. Compared with the already high CPU performance of a Xeon 6980P node, the GPUs of AMD MI300A and NVIDIA GH200 achieved additional speedups of 2.0-2.8 $\times$. 
Because the dominant kernels (FFT, GEMM, and eigenvalue solvers) are common to many plasma-fusion simulation workflows, the present  benchmark results are expected to benefit a broad range of simulation codes beyond RS-DFT.

Data availability: The source code of QUMASUN and the input files used for the present
benchmarks are openly available on GitHub:\\
\url{https://github.com/atsushi-m-ito/QUMASUN}

Software environment:
For the Xeon 6980P system, the Intel oneAPI compiler (version 2025.1) and Intel MKL were used.
On the MI300A system, HIP and ROCm version 6.4.3 were used.
On the GH200 system, CUDA version 13.0 was used.


The author thanks Dr.~Arimichi Takayama and Mr.~Yuto Toda for helpful discussions.
This work was supported by JSPS KAKENHI Grants JP23K17679, JP24H02251,
JP24K00617, and JP24H02246. Numerical calculations on the Xeon 6980P and
the MI300A were performed on the Plasma Simulator at QST-NIFS under the
auspices of the NIFS Collaboration Research Programs NIFS24KISM002 and
NIFS25KISM014. Numerical calculations on the GH200 were supported
by compute resources provided by NVIDIA. The author also used AI-based
language assistance (ChatGPT) for improving the clarity of the English
manuscript.


\end{document}